\documentclass{article}

   \usepackage[final,nonatbib]{neurips_2023_gaied}

\usepackage[utf8]{inputenc} % allow utf-8 input
\usepackage[T1]{fontenc}    % use 8-bit T1 fonts
\usepackage{hyperref}       % hyperlinks
\usepackage{url}            % simple URL typesetting
\usepackage{booktabs}       % professional-quality tables
\usepackage{amsfonts}       % blackboard math symbols
\usepackage{nicefrac}       % compact symbols for 1/2, etc.
\usepackage{microtype}      % microtypography
\usepackage{xcolor}         % colors
\usepackage{tabularx}
\usepackage{rotating}
\usepackage{arydshln}
\usepackage{multicol}
\usepackage{multirow}
\usepackage{listings}
\usepackage{array}
\usepackage{adjustbox}
\usepackage{mdframed}
\usepackage{longtable}
\usepackage{tcolorbox}
\usepackage{xcolor}
\usepackage{graphicx}
\usepackage{multirow}
\usepackage{listings}

\newmdenv[
  topline=false,
  bottomline=false,
  leftline=false,
  rightline=false,
  backgroundcolor=gray!10,
  font=\small,
]{codebox}

\definecolor{keywordcolor}{RGB}{0,0,255}
\definecolor{commentcolor}{RGB}{0,128,0}
\definecolor{stringcolor}{RGB}{255,0,0}

\newtcolorbox{boxK}{
    sharpish corners, % better drop shadow
    boxrule = 0pt,
    toprule = 4.5pt, % top rule weight
    enhanced,
    fuzzy shadow = {0pt}{-2pt}{-0.5pt}{0.5pt}{black!35} % {xshift}{yshift}{offset}{step}{options} 
}

\title{The Behavior of Large Language Models When Prompted to Generate Code Explanations}

\author{
  Priti Oli \thanks{Both authors contributed equally.}\\
  University of Memphis \\ 
  Memphis TN 38152,USA \\
  \texttt{poli@memphis.edu} \\
  \And
   Rabin Banjade \footnotemark[1]\\
  University of Memphis \\ 
  Memphis TN 38152,USA \\
  \texttt{rbnjade1@memphis.edu} \\
    \And
   Jeevan Chapagain\\
  University of Memphis \\ 
  Memphis TN 38152,USA \\
  \texttt{jchpgain@memphis.edu} \\
  \And
   Vasile Rus\\
  University of Memphis \\ 
  Memphis TN 38152,USA \\
  \texttt{vrus@memphis.edu} \\
}

\begin{document}

\maketitle

\begin{abstract}
This paper systematically investigates the generation of code explanations by Large Language Models (LLMs) for code examples commonly encountered in introductory programming courses. Our findings reveal significant variations in the nature of code explanations produced by LLMs, influenced by factors such as the wording of the prompt, the specific code examples under consideration, the programming language involved, the temperature parameter, and the version of the LLM. However, a consistent pattern emerges for Java and Python, where explanations exhibit a Flesch-Kincaid readability level of approximately 7-8 grade and a consistent lexical density, indicating the proportion of meaningful words relative to the total explanation size. Additionally, the generated explanations consistently achieve high scores for correctness, but lower scores on three other metrics: completeness, conciseness, and specificity.

\end{abstract}

\section{Introduction}
Explanation of code examples is an effective instructional tool to help students in intro-to-programming courses master programming concepts and develop other skills such as code comprehension skills \cite{garcesEtAl2019, oli2023improving}. Asking students to self-explain (a more active learning strategy) or simply reading worked-out examples accompanied by explanations provided by experts (a more passive learning strategy) was shown to have a positive impact on learning gains and code comprehension skills \cite{oli2023improving,tamang2021comparative}. Furthermore, interactive learning strategies such as scaffolded self-explanations while interacting with a human or computer tutor have also been explored with very positive results \cite{tamang2021comparative,oli2023improving}. Many types of explanations have been explored, such as `Explain-in-plain-English' ~\cite{murphy2012ability,murphy2012explain,whalley2006australasian}  or `stepwise explanation' \cite{sarsa22} or code comprehension-theory driven explanations ~\cite{tamang2021comparative}. In all those instances, experts are tasked to create corresponding explanations of the target code examples, which is tedious and expensive. Large Language Models (LLMs; \cite{gozalo2023}) have been recently explored as a potential solution to alleviate those authoring costs associated with expert-generated explanations. Automating the task of code explanation generation could lead to tremendous advantages in terms of scaling up the use of explanations across topics and domains. 

% Given the success of explanations for learning and the high costs of involving experts to generate the explanations, automating this task could lead to tremendous advantages in terms of scaling up the use of explanations across topics and domains. It should be noted that other natural language items are used in CS education contexts such as feedback and hints. While related to explanations, our focus, these other types of natural language items are different and serve different pedagogical purposes.

LLMs are impressive technologies relative to what has been done before in terms of content generation; however, there are several serious challenges with LLMs, such as non-deterministic behavior, generating incorrect output \cite{finnieansley22} or so-called hallucinogenic behavior (generating untrue facts, e.g., person X won award Y in year Z when in fact person W won that award), and data contamination (test data being very similar or identical to training data which can be considered some form of memorization; \cite{golchin2023time}). 
Those challenges suggest a more cautious approach to recommending or using such tools without proper, solid, systematic studies that can document the strengths and weaknesses of LLMs.
%For instance, the non-deterministic nature of LLMs leads to challenges with reproducing studies that may draw overly optimistic conclusions and make overly optimistic recommendations. 
Studies published recently with respect to generating code explanations (e.g., code summaries, line-by-line explanations; \cite{sarsa22,macneil22,mcneil23}) do not report important aspects of their use of LLMs, such as the actual prompts used or, if they explored a number of prompts and other parameters' values before finding the best combination of parameters that generated the kind of output wanted. Details of the prompt engineering and parameter space exploration process is not reported comprehensively. 

In response to these challenges and the need for a more systematic exploration of LLMs to better understand their behavior, we conduct a systematic exploration of the space of input parameters and an analysis of the behavior of LLMs focusing on the task of generating code explanations for code examples of the type used in intro-to-programming courses. Specifically, we study how five major input parameters alter the output of the LLMs, i.e., how the generated code explanations vary while those input parameters vary. We focused on the following five input parameters: input prompt wording, code example type, temperature, LLM model, and programming language.

%For instance, we conduct an analysis of LLMs for code explanation generation using both a deterministic and non-deterministic setting. The former is meant to allow other researchers to reproduce our results whereas the latter is meant to fully exploit the generative nature of LLMs. In general, the more generative/`creative' the LLMs (determined by a higher value of the temperature parameter) the more non-deterministic its output becomes. That is, there is a trade-off between generative power/`creativity' and determinism. 
%Furthermore, the more creative the more hallucinogenic the LLMs are, i.e., there is another trade-off between the correctness of output, i.e., fewer hallucinations, and `creativity', i.e., more likely to generate hallucinations if allowed to be as 'creative' as possible.

\section{Related Work}
% \input{related_work}
%We briefly review two major categories of related research. First, we review and emphasize important work with respect to evaluating LLMs. Second, we will look at prior related work that uses LLMs to generate code explanations. 

Relevant to our work, LLMs have been used and studied extensively so far from two major perspectives: (1) generating code from natural language descriptions (see NL2CODE survey paper for a good overview \cite{rogers2020, karmakar21, zan2023large}) and (2) generating explanations of code examples. Due to space reasons, we will focus on the latter.

%LLMs have become mainstream quite quickly due to their impressive ability and versatility to generate output/content from relatively simple inputs called prompts \cite{gozalobrizuela23}. LLMs can generate, for instance, code, images, and poetry, can do it in multiple languages like English, Spanish, or German, can solve exam problems, and so on. Nevertheless, LLMs have significant failures as well including reasoning (e.g., spatial, temporal, commonsense), math, coding, factual errors, and bias and discrimination \cite{borji23}. 

%LLMs including early transformer-based models such as BERT have been evaluated intensively in many cases for the purpose of really understanding in more depth how they really work \cite{rogers2020}. Some of those efforts focused on probing what LLMs really understand about code \cite{karmakar21}. Karmakar and Robbes \cite{karmakar21} used probing focusing on four aspects of code: surface, syntactic, structural, and semantic.

Sarsa and colleagues \cite{sarsa22} used LLMs to generate programming exercises (including sample solutions and test cases) and code explanations and assessed these items qualitatively and quantitatively. For the explanations part, they considered three types of explanations: a high-level description of the code, a problem statement-like description of the code, and a step-by-step explanation of the
code. However, their work focused on step-by-step explanations of code. They used this prompt, "Step-by-step explanation of the above program," and noted that this prompt `tended' to produce line-by-line explanations. The word `tended' hints at some inconsistent behavior, which our systematic study reported here confirms. They set the temperature to zero as they wanted `precise explanations instead of creative ones'. Sarsa and colleagues report 67\% correctness of the generated explanations and 90\% coverage, i.e., 90\% of `all parts of the code' were explained. They used in their study "a small set of exercises that have been featured in computing education research and that are often used in the teaching contexts of the researchers." On the one hand, using such widely used exercises lends more credibility and allows comparison to prior work; on the other hand, using well-known exercises for studying LLMs poses a major issue: data contamination.
%Indeed, a major problem of many studies using LLMs such as ChatGPT is data contamination, i.e., testing the LLMs on data they were trained on. The chances of the famous four exercises used by Sarsa and colleagues \cite{sarsa22} being used for training of the various versions of the GPT model are very high given the popularity of those exercises in various documents that are publicly accessible. 
Due to space reasons, we do not address the issue of data contamination in this paper but plan to explore it and report on it in future work.

%We will briefly address the issue of data contamination in our work presented here [VR: WE NEED TO USE A NON-FAMOUS EXERCISES THAT MINIMIZES DATA CONTAMINATION OR USED SOME WAY TO AVOID DATA CONTAMINATION].

McNeil and colleagues \cite{mcneil23} generated three types of code explanations using LLMs and integrated them into an interactive e-book used in a web software development course. The three types of explanations generated were a line-by-line explanation, a list of important concepts, and a high-level summary of the code. They report using the default parameter for GPT-3 Davinci model and, importantly, prompts sent to the two LLMs, GPT-3 and Codex, are different (shown in their paper in Figure 1): "Summarize and explain this code snippet" versus "Summarize and explain the goal of the above code." The latter only asks for a summary and explanation of the goal of the code. They claim, `Based on our preliminary findings, explanations appear helpful for learning'. Also, they report that ``at a cursory glance, we did not observe any significant mistakes, and the explanations were, in general, correct (although at times omitting details, as one would expect)." this is partly due to prompt engineering without providing details of the exact prompt engineering process. 
%They do raise the important question of whether LLMs' can be used for `live code explanations', i.e., on-demand.

MacNeil and colleagues used GPT-3 to generate `diverse ' explanations \cite{macneil22}. They explored what types of explanations GPT-3 can generate. Furthermore, the type of explanations was primed by the authors by the nature of the prompts they sent to the LLM. Notably, some of those prompts do not use the word `explanation' (or its morphological variations or closely related phrases such as `line-by-line explanation'). One example prompt they used is `Give a real analogy for this code.' Besides the fact that many of their prompts didn't use `explanation' keywords, the `diverse' term in the title can also be ambiguous in the sense that the diversity of explanations can be defined in very different ways. For instance, Maharjan and colleagues \cite{maharjan18} shows how linguistically diverse student explanations are, varying from just one word to sophisticated paraphrases of expert explanations.

%MacNeil and colleagues elicit code descriptions that talk about time complexity. Code explanations that are meant to help students learn and comprehend code do not typically focus on time complexity ~\cite{oli2023improving}. In fact, major comprehension theories do not emphasize such aspects. Rather, they focus on students' being able to construct an accurate mental model of the structure and functionality of the code in order to predict its behavior and output. Ideally, complexity should be addressed but typically it is not in intro-to-programming courses, maybe, with a few exceptions depending on the actual topic or instructor.
In sum, prior work on using LLMs to generate code explanations is limited in several ways: they have a wide range of views of what a code explanation should be, they do not always report the exact prompts they used, they do not report other important parameters, or do not vary such parameters systematically (e.g., they use the default value of the temperature parameter), nor analyze how such other important factors may alter the nature of the explanations generated.

%It is not our intent to explore the entire space of all the factors that may affect the generated code explanations. The main point of our work is to illustrate that the nature of the generated explanations can vary vastly and that future work reporting explanations generated by LLMs should detail all the factors, i.e., specify the point in the space of possible values of all the important factors. At the minimum, future studies should report the exact prompt and the prompt re-engineering process.
%, the temperature parameter, and how they filtered the LLMs' output in particular with respect to detecting hallucinations.

\section{Methodology}

We detail in this section the methodology we used to systematically explore the space of input parameters for LLMs and the prompt engineering process.
%We followed a methodology that emphasizes a systematic study of LLMs' behavior/output based on a controlled change of the input across a number of dimensions: input prompt, temperature, code example, LLM, and programming language. For instance, when looking at how LLMs behave along the programming language dimension, we keep all other input factors the same, e.g., using the same programming example with the same input prompt and the same temperature, i.e., we vary the programming language in which the code example is written.

\subsection{Code Explanation Generation Across various factors}

Prompting LLMs is probably the most important factor, as this is the input based on each, the LLMs will produce the output; it is also the trickiest factor as the input prompt can alter the behavior/output of the LLMs substantially, e.g., prompts can be very simple requests such as P1 in Table \ref{tab:prompt_used} {\em Can you explain this code?} to heavily contextualized prompts (see prompts C1 and C2 in the table) where the user is providing a rich paragraph of contextualized information such as {\em You are a tutor who is supposed to teach students programming. The students are novices ... } (see the table for the full prompt) to iterative prompts to a few shot prompting in which examples of code and explanations are given to the LLMs to self-prompts, i.e., LLM-generated prompts - the user creates a simple prompt asking the LLM to generate a prompt for LLMs which in turn will be used to prompt the LLM.

This paper focuses on simple and contextualized prompts such as those shown in Table \ref{tab:prompt_used} and how their wording influences the LLMs' behavior. For the wording, we started with simple wording inspired by previous work (see \cite{mcneil23}) and then made small, step-wise changes to the wording while also asking for different types of explanations based on code comprehension theories~\cite{schulte2010introduction} (e.g., generic explain the code, line-by-line, block-level, summary, contextualized, used-by-others).  
%That is, we analyze how the wording/phrasing of the prompt, which can vary substantially, alters the LLMs' behavior. Exploring LLMs' behavior across all the categories of prompts mentioned above is beyond the scope of this paper. We will explore it as part of our future work.

\begin{table}[]
    \centering
    % \begin{tabular}{c|l}
  \small
    \begin{tabularx}{\textwidth}{c|p{9cm}|p{2.5cm}}

       \hline
        Symbol & Prompt & Comments  \\
        \hline
        P1 & Can you explain this code? & \multirow{3}{*
        }{Simple prompts and }  \\
        % & & and variations \\
        P2 & Can you self-explain this code? & \\
        P3 & Can you explain this code to a learner? & variation \\
        P4 & Can you explain this code to someone learning to program? &  \\ 
        \hline
        P5 & Can you summarize this code? & \multirow{2}{*}{Prompt for summary}    \\
        P6 & Can you summarize this code for a learner? &  \\ 
        \hline
%        \hline
        P7 & Can you explain this code at statement level? & line by line explanations \\
        \hline
        P8 & Can you explain this code at block level? & logical/functional  \\
        P9 & Can you explain this code without breaking down individual statements? & level explanations \\

        \hline
%         P10 & What is the complexity of this code and why? & ?????? \\
%        P11 & Can you explain this code without code complexity? & ?????? \\
%        P12 & Can you explain this code without breaking down individual statements and without code complexity? & ??????\\
        % \hline
         % & & \multirow{2}{*}{Contextualized}  \\
         C1 & \textbf{Context:} You are a tutor who is supposed to teach students programming. The students are novices and your task is to give students learning tasks. One of the learning tasks is to read and explain code examples. The explanation should clearly articulate what the goal of the code is, what the major functional blocks are, and implementation details.

        \textbf{Prompt:} Given this Java code, explain the code to your students in order to help them understand what the code does and learn the covered programming concepts. 
  & \multirow{2}{*}{Contextualized}  \\
    C2 &\textbf{ Context:} You are supposed to read code examples in order to understand them. You will be given one code example at a time, your task is to read each code example as carefully as you can and then explain your understanding of the code as best as you can.

        \textbf{Prompt:} Given this Java code, read the code carefully and explain what it does to potential students who learn programming. & \\

        \hline
        
        D1 & Explain the following code line by line as a bulleted list: & Prompts used in \\
        D2 &  Give a detailed explanation of the purpose of the following code: & prior work~\cite{mcneil23} \\
        D3 & Summarize and explain the goal of the above code & ~ \\
         \hline
    \end{tabularx}
    \caption{\small{The input prompts used from simple to contextualized.}}
    \label{tab:prompt_used}
\end{table}

In addition to prompt wording, we investigate four factors influencing code explanations: language model, temperature, code example complexity (basic, intermediate, advanced), and programming language (Java, Python, C++). The language models we investigated were OpenAI's ChatGPT-3.5-turbo-0613 and chatGPT-4-0613 ~\cite{openai2023gpt4}, and Meta's open-source model LLMa2-chat\footnote{https://huggingface.co/meta-llama/Llama-2-7b-chat-hf}~\cite{llama2}). The temperature parameter can take values between 0 and 2, with values closer to 0, limiting the diversity of the output by limiting how `generative/creative' the LLM is. 
% This multi-model approach allowed us to tailor our output to specific use cases, providing a nuanced perspective on the prompts under investigation and enabling us to draw meaningful insights from the generated text.

All our code examples are typical examples used in intro-to-programming courses (CS1 and CS2). We chose several examples from very basic to intermediate to more complex. Specifically, we used the following code examples: \textit{ calculate the average of numbers, calculate the area of a circle, search a number using binary search, translate a point, and generate a bingo board}, the Java version of which is shown in Appendix~\ref{app:code_examples}. The code examples vary based on concepts, difficulty, and structure.

%We use three relatively popular programming languages: Python, Java, and C++. It is well known that ChatGPT, for instance, works well with Python. We explored to what extent the generated code explanations varied across programming languages while keeping other factors constant.

\subsection{The Data: The Generated Code Explanations}
We have generated explanations using 13 input prompts, 3 LLMs, 5 different temperature levels (0, 0.5, 1, 1.5, 2.0), 5 code examples, and 3 programming languages for a total of 3,510 code explanations. Explanations with temperature value 2 were discarded in our analysis as they are non-sensical for all 3 LLMs. We set a maximum token limit of 500 for the generated explanations.

%\subsection{The Data: The Generated Set of Code Explanations}

\section{Results}

We analyzed the generated explanations using a number of criteria, including readability, lexical diversity, and correctness, as well as in terms of their nature. To conduct the analysis, we extracted a subset of 200 code explanations through stratified sampling from the set of 3,510 explanations except for the ones for temperature 2.0.

% This section summarizes and describes the most interesting results of our systematic study of the behavior of LLMs when varying the five factors described earlier: prompt, temperature, code example type, programming language, and LLM version. We conducted both a qualitative and quantitative analysis of the LLMs' output.

\subsection{Quantitative Analysis}

The quantitative analysis looked at some surface-level properties of the generated explanations, such as length in terms of the number of words, number of sentences, readability, lexical density, and vocabulary of the generated explanations. 
\textit{Lexical diversity} is the range of
variety of unique tokens or vocabulary used within a specific text. \textit{Lexical density} refers to the measure of content words used in a text, including nouns, adjectives, verbs, and adverbs, which collectively contribute to the overall meaning of the text. It is calculated by dividing the number of content words by total number of tokens. For readability, we report Flesch–Kincaid reading grade level~\cite{kincaid1975derivation}
 that assesses the ease of understanding a passage in English based on sentence length and word complexity.

Table \ref{quantitative_eval_short} presents metrics explanations generated for different prompts, including lexical diversity (Vocabulary), the average number of tokens, and the number of sentences. Vocabulary scores vary slightly, with some prompts having a more diverse vocabulary than others. The average number of tokens ranges widely, with some explanations for some prompts being considerably longer than others. One interesting observation is that explanations generated from prompts C1 and D2 have exceptionally high average numbers of tokens, making them significantly longer than the other prompts. This is not surprising as C1 is asking the LLM to act like a tutor teaching novices, and thus, detailed explanations are needed, whereas D2 explicitly asks for detailed explanations. Readability level is probably the single most consistent feature of the generated explanations hovering around 60 indicating a 7-grade level for Java and Python but varies widely for C++ (see details in the Appendix ~\ref{appendix_quant_analysis}). The other consistent feature is lexical density. Interestingly, as LLM generates longer explanations in response to certain prompts or certain values of other input parameters, such as the type of code example, the vocabulary or lexical diversity of those explanations also increases, keeping the lexical density relatively constant at around 0.43-0.45. This is true across all three LLMs we investigated (see details in the Appendix ~\ref{appendix_quant_analysis} ).

\begin{table}[h!]
   
\label{quantitative_eval_short}
    \centering
    \caption{\small{Quantitative Evaluation Scores Across various Prompts}}
    \small
\begin{tabular}{lrrrrrrr}
\toprule

prompts &   Rdb &  LexDensity &  Vocabulary &  Tokens(mean) &   Tokens(sd) &   Sent(M) &  Sent(sd) \\
\midrule  
     P1 & 62.00 &    0.43 &      170.33 &        278.67 &        74.67 &     26.00 &      8.33 \\
     P2 & 63.33 &    0.43 &      165.00 &        271.67 &        84.67 &     25.33 &      8.67 \\
     P3 & 68.33 &    0.40 &      109.33 &        152.67 &        67.00 &     11.67 &      6.67 \\
     P4 & 63.67 &    0.43 &      183.33 &        300.67 &        92.00 &     29.00 &      9.67 \\
     P5 & 61.67 &    0.43 &      193.33 &        316.33 &        88.67 &     28.00 &      9.33 \\
     P6 & 63.67 &    0.47 &      118.00 &        197.67 &        66.00 &     16.67 &      6.33 \\
     P7 & 61.67 &    0.43 &      190.67 &        302.00 &        76.33 &     27.67 &      9.00 \\
     P8 & 62.33 &    0.47 &      183.67 &        295.67 &        75.67 &     26.67 &      8.33 \\
     P9 & 62.00 &    0.50 &      109.67 &        182.00 &        78.33 &     15.00 &      8.00 \\
     C1 & 60.67 &    0.50 &      231.00 &        380.33 &       110.67 &     29.67 &     11.67 \\
     C2 & 62.00 &    0.50 &      169.00 &        279.00 &        82.33 &     21.00 &      8.00 \\
     D1 & 60.00 &    0.47 &      206.67 &        327.33 &       104.33 &     25.33 &     10.00 \\
     D2 & 59.00 &    0.50 &      198.00 &        327.33 &       100.33 &     27.00 &     10.33 \\
     D3 & 60.33 &    0.50 &      130.33 &        216.33 &        84.33 &     16.00 &      8.00 \\
\bottomrule
\end{tabular}
\label{quantitative_eval}
\end{table}

\subsection{Qualitative Analysis}
We examined the quality of the generated explanation using the following criteria: accuracy, completeness, conciseness, and specificity (see Table \ref{table_qualitive_result}). The first three criteria were initially used by Sridhara et al ~\cite{sridhara2010towards} to evaluate automatically generated code summaries and have been the common benchmark in various studies ~\cite{su2023distilled}.

Accuracy indicates whether the explanations are correct. Completeness evaluates whether the explanations contain all essential information for understanding the code. Conciseness examined whether the explanations were free from excessive or unnecessary information. Specificity determined if the explanations were tailored to the specific code examples provided without focusing only on the underlying algorithm.

Two graduate students independently evaluated these explanations using the predefined qualitative metrics on a binary scale, and their assessments showed substantial agreement, as reflected by high Cohen's Kappa coefficients  with scores of 0.957 for correctness, 0.965 for completeness, 0.935 for concision, and 0.923 for specificity. For LLAMA2 and ChatGPT-3.5 Turbo, we considered examples with temperature parameters not exceeding 1.5, while for chatGPT-4, we limited the temperature to 1 due to observations that higher temperatures led to non-sensical output in most cases.

\begin{table}[htbp]
    \centering
    \caption{\small{Qualitative Evaluation Scores Across various factors}}
    \small
    \begin{tabular}{llcccc}
        \toprule
        Factors & Values & Correctness & Completeness & Concision & Specificity \\
        \hline
       \multirow{14}{*}{Prompts} 
        & P1 &  0.92 & 0.92 & 0.54 & 0.85 \\
        & P2 &  0.85 & 1.00 & 0.62 & 1.00 \\
        & P3 &  0.86 & 0.79 & 0.57 & 0.64 \\
        & P4 &  1.00 & 0.92 & 0.42 & 0.88 \\
        & P5 &  0.86 & 0.43 & 0.93 & 0.57 \\
        & P6 &  1.00 & 0.86 & 0.93 & 0.79 \\
        & P7 &  1.00 & 0.86 & 0.39 & 0.86 \\
        & P8 &  1.00 & 0.91 & 0.36 & 0.82 \\
        & P9 &  1.00 & 0.60 & 0.67 & 0.47 \\
        & C1 &  0.86 & 0.86 & 0.43 & 0.71 \\
        & C2 &  1.00 & 0.92 & 0.62 & 1.00 \\
        & D1 &  0.86 & 0.79 & 0.64 & 0.75 \\
        & D2 &  0.96 & 1.00 & 0.46 & 1.00 \\
        & D3 &  1.00 & 0.83 & 0.62 & 0.88 \\
        % \hdashline
        % & Avg & 0.92 & 0.84 & 0.62 & 0.80\\
        \hline
        \multirow{3}{*}{Temperature} & 0 &  0.98 & 0.81 & 0.62 & 0.84 \\ 
        & 0.5 &  0.96 & 0.91 & 0.56 & 0.81 \\ 
        & 1 &  0.93 & 0.84 & 0.60 & 0.82 \\ 
        & 1.5 &  0.71 & 0.44 & 0.41 & 0.35 \\ 
        % \hdashline
        % & Avg & 0.89 & 0.76 & 0.56 & 0.69\\
        
        \hline
        \multirow{3}{*}{Model} 
        & gpt-3.5-turbo  &  0.97 & 0.81 & 0.75 & 0.82 \\
        &  gpt-4  &  0.96 & 0.82 & 0.56 & 0.88 \\
        & llama2  &  0.88 & 0.82 & 0.41 & 0.66 \\
        % \hdashline
        % & Avg & 0.93 & 0.81 & 0.57 & 0.78\\
        
        \hline

        \multirow{3}{*}{Language} 
        & Java  &  0.95 & 0.80 & 0.66 & 0.80 \\
        & Python  &  0.91 & 0.78 & 0.43 & 0.83 \\
        & CPP  &  0.95 & 0.87 & 0.66 & 0.71 \\
        % \hdashline
        % & Avg & 0.93 & 0.81 & 0.58 & 0.78\\
        \hline
        \multirow{5}{*}{Code Examples} 
        & AreaOf Circle  &  0.91 & 0.89 & 0.49 & 0.92 \\
        & AvgOfNumbers  &  0.96 & 0.80 & 0.50 & 1.00 \\
        & Point  &  0.94 & 0.83 & 0.59 & 0.81 \\
        & BingoBoard  &  0.92 & 0.83 & 0.83 & 0.80 \\
        & BinarySearch  &  0.96 & 0.77 & 0.67 & 0.66 \\
        % \hdashline
        % & Avg & 0.93 & 0.82 & 0.58 & 0.78\\
        \hline
        Average &  & 0.93 & 0.82 & 0.58 & 0.77\\
        \hline
        
    \end{tabular}
    \label{table_qualitive_result}
\end{table}

The accuracy of the generated explanations was measured at 93\%, indicating a generally high level of correctness. However, only 82\% of the explanations were deemed complete, implying room for improvement in providing comprehensive information. Moreover, conciseness was a concern, with just 58\% of the explanations considered concise. Furthermore, only 77\%  of the explanations were effectively specific, focusing on the provided code example without excessively delving into underlying algorithmic details.

\subsection{Diversity or Inconsistency}

%A potential use of LLM-generated code explanations is by developers of learning technologies serving two purposes: explanations for worked-out examples and explanations for assessing students' self-explanations. The former may require just one good explanation of a given target code. In contrast, the latter requires a set of alternative explanations as a way to mitigate the diversity of explanations that students' explanations exhibit \cite{banjade16,maharjan18,khayi20}.

%Probably, the most consistent feature of the generated explanations is the readability level which is extremely similar across prompts and models as can be seen in Table \ref{tab:Variation_prompt2}. The readability level is hovering around 70 which is equivalent to 7-8 grade level. The only exception is for explanations generated by GPT-4 when using the three prompts used by others in the past. The other consistent feature is lexical density. Interestingly, as LLM generate longer explanations in response to certain prompts or certain values of other input parameters such as the type of code example, the vocabulary or lexical diversity of those explanations also increases keeping the lexical density relatively constant at around 0.43-0.45. This is true across all three LLMs that we investigated.

As expected, the generated explanations vary considerably as the input prompt's wording varies. First, the different wordings ask for different kinds of explanations. To some degree, but not always, the LLMs somehow generate what the prompt intended, e.g., a summary, a line-by-line explanation, or more sophisticated explanations as those based on code comprehension theories \cite{tamang2021comparative}. However, many other times it does not. For instance, GPT4 does generate sometimes what is being asked by the contextualized prompt C1, which asks for the goal, functional blocks, and implementation details. For this prompt, ChatGPT 3.5 and Llama do not follow the prompt's instructions. One major failure is the prompt asking for block-level explanations. All LLMs we tried cannot properly handle this prompt with few exceptions. The qualitative evaluation per prompt category is shown in the Table \ref{table_qualitive_result}. 

Second, when varying the input prompt's wording, adding a single word, e.g., {\em learner}, can lead to significant differences in the generated explanations. For instance, the difference between prompts P1 and P3, on the one hand, and P5 and P6, on the other hand, is just the addition of suggesting to explain to a learner. This relatively small change leads to significantly larger explanations (see Table \ref{tab:Variation_prompt1}) in terms of their overall size (number of tokens/words) as well as vocabulary size (unique tokens/words). This is the case for GPT3.5 and GPT 4.0 but not so much for Llama. The default LLama behavior is to add conversational, code-irrelevant text to the generated explanation such as {\em Hello students!}. We tried to suppress this conversational default style, which we managed to a great extent (for instance we managed to eliminate USER and ASSISTANT standard dialogue structure in the Llama output), but the Llama model still includes conversational snippets in the output.

There are also major differences/inconsistencies among the generated explanations in terms of their general structure. For instance, for the same prompt, the nature of the generated explanations varies when varying the type of code examples. Sometimes, the explanation starts with a short code summary followed by a breakdown of the code. Other times, it does not follow this general pattern. One explanation could be that various code examples, particularly those used in intro-to-programming courses, have different types of explanations available as training instances in publicly accessible resources such as websites and textbooks thus leading to different types of LLM-generated explanations. Consider variation in code example generation for three different programs by GPT as shown in Appendix ~\ref{app:Variation among explanations}

%WE NEED THREE EXAMPLES IN THE APPENDIX  (appendix ~\ref{app:Variation among explanations})

\section{Conclusion and Future Work}

Some of the major lessons learned from our study are: LLMs have a tendency to generate longer explanations, and when the temperature is greater than 1 the output is not very useful (non-sensical, more or less). GPT-4 works, in general, better than ChatGPT 3.5 and Llama2. They all generate widely different types of explanations depending on the actual wording of the input prompt, temperature parameter, and code example type. Overall, all LLMs show a great deal of diversity, which can be seen as a form of inconsistency as differences in the generated explanations become very wide. There are three major consequences of this diversity/inconsistency: (1) the exact parameters used by researchers to generate code explanations must be well documented as otherwise their work cannot be used or replicated; (2) it is challenging to use LLMs' diverse/inconsistent output for certain pedagogical needs that are supposed to rely on explanations of code that follow a particular theory without additional work; and (3) LLMs may be used to obtain a set of rough explanations, a substantial human effort would be needed to refine those explanations.

A comprehensive understanding of how various factors influence LLM behavior in code explanation, including prompt creation, is essential to establish clear guidelines for using these technologies in education. To facilitate the use of LLMs in generating code explanations, a repository of prompts with user ratings and parameters like temperature should be established.
%Furthermore, considering the high likelihood of data contamination in widely used code examples and the availability of reliable explanations from non-hallucinatory sources on the web, the true value and drawbacks of LLMs in this context remain unclear. 

%Prompt engineering, although time-consuming, may still yield unpredictable outputs. For this reason, to facilitate the use of LLMs in generating code explanations, a repository of prompts with user ratings and parameters like temperature should be established. This resource would provide a valuable starting point for those seeking to employ LLMs for code explanations, whether in offline or online settings, and mitigate the challenges of searching for effective prompts.
There is significant work ahead to fully comprehend the generation and utilization of LLM-generated explanations. This research represents a step in that direction, with plans for continued exploration of this topic.

\section*{Acknowledgements}

This  work  has  been  supported  by  the following grants awarded to Dr. Vasile Rus: the Learner Data Institute (NSF award 1934745); CSEdPad (NSF award 1822816); iCODE (IES award R305A220385). The opinions, findings, and results are solely those of the authors and do not reflect those of NSF or IES.

% NeurIPS allows any reference format 
\bibliographystyle{acm}
\bibliography{bibliography}

%This bibliography style gets last name and first name but references number collides with line number of the page on second half of the references
% \bibliographystyle{splncs04}  

% VR on 9/23/2023: I commented the pagebreak below as it makes the references weird on the last page of references if they are too few.
% \pagebreak
\clearpage
\section{Appendix}
\subsection{Appendix A: Code Examples }
\label{app:code_examples}

1. AreaOfCircle.java

\begin{lstlisting}
public class AreaOfCircle {
    public static void main(String[] args) {
        final double PI = 3.14159;
        double radius = 5.8;
        double area = radius * radius * PI;
        System.out.printf("The area of the circle is %.2f\n", area);
    }
}
\end{lstlisting}

2. AverageOfNumbers.java
\begin{lstlisting}
public class AverageOfNumbers {
   public static void main(String[] args) {
      
       double[] numArray = {8,6,11,7};
       double sum = 0.0;
       double average;
      
       for (int i = 0; i < numArray.length; i++) {        
           sum += numArray[i];
       }
    
       average = sum / numArray.length;
       System.out.format("The average is: %.2f", average);

   }
}
\end{lstlisting}

3. BinarySearch.java

\begin{lstlisting}
public class BinarySearch {
    public static int binarySearch(int[] arr, int x) {
        int left = 0;
        int right = arr.length - 1;

        while (left <= right) {
            int mid = (left + right) / 2;
            if (arr[mid] == x) {
                return mid;
            } else if (arr[mid] < x) {
                left = mid + 1;
            } else {
                right = mid - 1;
            }
        }
        return -1;
    }

    public static void main(String a[]) {
        int[] luckyNumbers = {1, 3, 5, 7, 9};
        int guess = 5;
        int index = binarySearch(luckyNumbers, guess);

        if (index == -1) {
            System.out.println("Value not found, "
                + "the guessed number is not a lucky number");
        } else {
            System.out.println("Value found, " 
                + "the number is a lucky number");
        }

    }
}
\end{lstlisting}

4. BingoBoard.java

\begin{lstlisting}
import java.util.Random; 
public class twoDimensionalArraysBingoBoard {
   public static void main(String[] args) {
		
		int[][] bingoBoard = new int[5][5];
		Random rand = new Random();
		
		for ( int i = 0 ; i < 5 ; i++ )
		{
			for ( int j = 0 ; j < 5 ; j++ )
			{				
				while ( (bingoBoard[i][j] = rand.nextInt (75)) == 0) ;			
				System.out.print( "board square [" + i + ", " + j + "]" +
                " = " + bingoBoard[i][j] + "\n") ;
			}
			System.out.println( "" ) ;
		}
		
		for ( int i = 0 ; i < 5 ; i++ )
		{
			for ( int j = 0 ; j < 5 ; j++ )
			{
				System.out.print(bingoBoard[i][j] + " ") ;
			}
			System.out.println( "" ) ;
		}
   }
}


\end{lstlisting}

5. Point.java
\begin{lstlisting}
public class Point {

    private int x;
    private int y;


    public void setX(int newX) {
        this.x = newX;
    }
    public int getX() {
        return x;
    }

    public void setY(int newY) {
        this.y = newY;
    }
    public int getY() {
        return y;
    }

    public void translate(int dx, int dy) {
        x += dx;
        y += dy;
    }

    public static void main(String[] args) {

        Point point = new Point();
        point.setX(7);
        point.setY(2);


        point.translate(11, 6);
        System.out.println("The new coordinates: (" 
         + point.getX() + ", " + point.getY() + ")");

        }
    }
\end{lstlisting}

\subsection{Appendix B: Variation of Generated Explanations across code examples for GPT-4, Prompt : C1, Temperature 0}
\label{app:Variation among explanations}
% \textbf{Temperature}:0 \\
% \textbf{Model:} ChatGPT-4.0 \\
% \textbf{Prompt:}C1 \\

\begin{tcolorbox}

\textbf{AreaOfCircle.java}\\

This Java program calculates and prints the area of a circle with a given radius. 

The program is divided into several parts:

1. \textbf{Class Declaration:} The program begins with the declaration of a public class named `AreaOfCircle'. In Java, every application must contain a main class that wraps up all the program code.

2. \textbf{Main Method:} Inside the class, the `main' method is defined. This is the entry point for any Java application. The Java Virtual Machine (JVM) calls the main method when the program starts. The `String[] args' parameter represents command line arguments.

3. \textbf{Constant Declaration:} Inside the main method, a constant `PI' is declared with the `final' keyword and initialized with the value `3.14159'. The `final' keyword in Java is used to indicate that a variable is a constant - its value cannot be changed once it is initialized.

4. \textbf{Variable Declaration:} Two double-type variables `radius' and `area' are declared. The `radius' variable is initialized with the value `5.8', and `area' is not yet initialized.

5. \textbf{Area Calculation:} The area of the circle is calculated using the formula `area = radius * radius * PI;' and the result is stored in the `area' variable.

6. \textbf{Output:} The `System.out.printf' function is used to print the result. The `\%.2f' is a format specifier, it means we are formatting the floating point number up to 2 decimal places.

The major functional blocks of this program are the main method, the area calculation, and the output. The program uses basic concepts such as variable and constant declaration, arithmetic operations, and output formatting.

\end{tcolorbox}

\begin{tcolorbox}

\textbf{BinarySearch.java}\\

This Java code is an implementation of the Binary Search algorithm. The purpose of this code is to find a specific number (in this case, the variable `guess') in a sorted array of numbers (in this case, the array `luckyNumbers'). 

The major functional blocks of this code are:

\begin{enumerate}
  \item The binarySearch method: This method takes in an array and a number as parameters. It uses the Binary Search algorithm to find the number in the array. The Binary Search algorithm works by repeatedly dividing the array into two halves until the target value is found. If the target value matches the middle element, its position in the array is returned. If the target value is less or more than the middle element, the search continues on the lower or upper half respectively. If the target value is not found after checking all elements, the method returns -1.

  \item The main method: This is the entry point of the program. It initializes an array of `luckyNumbers' and a `guess' number. It then calls the binarySearch method with these values. Depending on the returned value from the binarySearch method, it prints out whether the guessed number is a lucky number or not.
\end{enumerate}

In terms of implementation details, the binarySearch method uses a while loop to repeatedly check the middle element of the array. The `left' and `right' variables define the current segment of the array that is being searched. The `mid' variable is calculated as the average of `left' and `right'. Depending on the comparison of `arr[mid]' and `x', either the `left' or `right' boundary is adjusted, effectively halving the search space for the next iteration. This halving of the search space is what makes Binary Search a very efficient algorithm for large data sets.

\end{tcolorbox}

\begin{tcolorbox}

\textbf{Point.java}\\

This Java code defines a class named `Point' that represents a point in a two-dimensional space. The class has two private instance variables `x' and `y' that represent the x-coordinate and y-coordinate of the point, respectively.

The class provides getter and setter methods for both `x` and `y'. The `setX(int newX)' and `setY(int newY)' methods are used to set the values of `x' and `y', respectively. The `getX()' and `getY()' methods are used to retrieve the current values of `x' and `y', respectively.

The `translate(int dx, int dy)' method is used to change the position of the point by a certain amount. The parameters `dx' and `dy' represent the amount to change the x-coordinate and y-coordinate, respectively. The method adds `dx' to `x' and `dy' to `y', effectively moving the point to a new position.

The `main(String[] args)' method is the entry point of the program. It creates a new `Point' object, sets its `x' and `y' values to 7 and 2, respectively, then translates the point by 11 in the x-direction and 6 in the y-direction. Finally, it prints out the new coordinates of the point.

This code is a good example of encapsulation in object-oriented programming. The `x' and `y' variables are hidden from other classes and can only be accessed through the getter and setter methods. This ensures that the internal state of the `Point' object is always controlled and can't be changed arbitrarily by other parts of the program.

\end{tcolorbox}

\subsection{Appendix C: The Behavior of Explanations Generated by LLM }
\label{app:explanation-code-example}
% good explanation
\begin{tcolorbox}
\textbf{Code Example}: BinarySearch.java \\
\textbf{Large Language Model}: GPT-4.0 \\
\textbf{Temperature}: 1\\
\textbf{Prompt}: P2:Can you self-explain this code?
\\

\textbf{Generated Explanation}:

This is a Java program that performs a binary search on an array. This kind of search is efficient when working with sorted data, because it essentially divides the array in half with every iteration, reducing the amount of data to be searched with each step. 
The class is called `BinarySearch' and contains two methods: `binarySearch()' and `main()'.
In the `binarySearch()' method:
- It takes an array (`arr') and a number (`x') as input.
- It initializes `left' and `right' pointers, pointing to the start and end of the array respectively.
- It goes into a loop where it continues to divide the array until the array is empty (i.e., the left pointer becomes greater than the right pointer).
    - It calculates a mid-point index, and checks whether the element at the mid-point is equal to the input number `x'. If equal, it returns the mid-point.
    - If the number at mid-point is less than `x', that means `x' could be in the right half of the array, so it shifts the `left' pointer to the next element after the mid-point.
    - If the number at mid-point is greater than `x', that means `x' could be in the left half of the array, so it shifts the `right' pointer to the previous element before the mid-point. 
- If it has searched through the entire array and hasn't found `x', it returns -1.
The `main()' method:
- It creates an array called `luckyNumbers'.
- It does a binary search for a guessed number (`guess') in `luckyNumbers'.
- If the binary search returns an index of -1, it means the number wasn't found, so it prints "Value not found, the guessed number is not lucky number".
- Otherwise, it prints "Value found the number is lucky number".
This program relies methods in array sorting and binary search algorithm to quickly search for data in a pre-sorted array.
\\

\textbf{Comment}:
This is an example of good explanation generated by the LLM. The explanation is correct, complete, concise and specific i.e the code is well-tailored to the context in the code that checks whether the guessed number is lucky number or not. 
\end{tcolorbox}

\begin{tcolorbox}
\textbf{Code Example}: AreaOfCircle.java \\
\textbf{Large Language Model}: Llama2 \\
\textbf{Temperature}:1 \\
\textbf{Prompt}: P8:Can you explain this code at block level?
\\ 

\textbf{Generated Explanation}:

The code starts by defining a constant `PI' to be equal to the value 3.14159, which is the value of the mathematical constant pi.  \textcolor{blue}{ Next, the code prompts the user to enter the radius of the circle. This value is stored in the variable `radius'.}
 
 Then, the code calculates the area of the circle using the formula `area = radius * radius * PI'. This calculation is performed using the `* ' operator, which performs multiplication.
 
 Finally, the code prints the result of the calculation to the console using the `System.out.printf()' method. This method takes two arguments: the first is a format string, which in this case is `\%.2f', which specifies that the output should be a floating-point number with two decimal places. The second argument is the value to be printed, which is the calculated area of the circle.
 
 Overall, this code provides a simple way to calculate the area of a circle using the formula `area = radius * radius * PI', and it allows the user to easily input the radius of the circle." 
 \newline

\textbf{Comment}:

The explanation erroneously mentions that 
"the code prompts..." when, in fact, there is no user input or prompt involved. Instead, the value of the double variable radius is explicitly set to 5.8 within the code.

\end{tcolorbox}

\begin{tcolorbox}
\textbf{Code Example}: AreaOfCircle.java \\
\textbf{Large Language Model}: GPT-4 \\
\textbf{Temperature}:0.5 \\
\textbf{Prompt}: C2
\\ 

\textbf{Generated Explanation}:
"This Java code calculates and prints the area of a circle with a given radius. Here is a breakdown of what each part of the code does:

- `public class AreaOfCircle': This line declares a public class named ""AreaOfCircle"". \textcolor{blue}{ A class is a blueprint for creating objects in Java.}

- `public static void main(String[] args)': This line defines the main method.  \textcolor{blue}{This is the entry point for any Java program. The Java Virtual Machine (JVM) calls the main method when the program starts.}

- `final double PI = 3.14159;': This line declares a constant named ""PI"" with a value of 3.14159. The keyword ""final"" is used to declare a constant, meaning its value cannot be changed once assigned.

- `double radius = 5.8;': This line declares a variable named ""radius"" of type double (which can hold decimal values) and assigns it a value of 5.8.

- `double area;': This line declares a variable named ""area"" of type double, but doesn't assign it a value yet.

\texttt{area = radius * radius * $\pi$;} : This line calculates the area of the circle using the formula $\texttt{area} = \pi r^2$ (pi times the radius squared), and assigns the result to the \texttt{area} variable.

- `System.out.printf(""The area for the circle of radius \%.2f is \%.2f "", radius, area);': This line prints a formatted string to the console. The ""\%.2f"" is a placeholder for a floating-point number with 2 decimal places.  \textcolor{blue}{The first \%.2f will be replaced by the value of ""radius"", and the second \%.2f will be replaced by the value of ""area"".}

So, if you run this program, it will print ""The area for the circle of radius 5.80 is 105.68 "", which is the area of a circle with a radius of 5.8 units."
\\
\textbf{Comment}:

The explanation, while correct, complete and specific lacks concision.

\end{tcolorbox}

% LLM does not provide contextualized explanation

\begin{tcolorbox}
\textbf{Code Example}: BinarySearch.java \\
\textbf{Large Language Model}: GPT-3.5\\
\textbf{Temperature}:0.5 \\
\textbf{Prompt}: P2:Can you self-explain this code?
\\ 

\textbf{Generated Explanation}:

This code implements the binary search algorithm to search for a given value in an array. 
The `binarySearch' method takes in an array `arr' and a value `x' to search for. It initializes two variables `left' and `right' to represent the left and right boundaries of the search range, respectively. 
The method then enters a while loop that continues as long as `left' is less than or equal to `right'. Inside the loop, it calculates the middle index of the search range using the formula `(left + right) / 2'. It compares the value at the middle index `arr[mid]' with the target value `x'. If they are equal, it returns the index `mid'. If `arr[mid]' is less than `x', it updates `left' to `mid + 1' to search the right half of the array. Otherwise, it updates `right' to `mid - 1' to search the left half of the array. 
If the while loop terminates without finding the target value, the method returns -1 to indicate that the value was not found in the array. 
In the `main` method, an array `luckyNumbers' is initialized with some values and a target value `guess' is set to 5. The `binarySearch' method is called with `luckyNumbers' and `guess' as arguments, and the returned index is stored in the `index' variable. 
 \textcolor{blue}{
The code then checks if `index' is -1. If it is, it prints a message indicating that the value was not found. Otherwise, it prints a message indicating that the value was found}
\\

\textbf{Comment}:
The LLM generated explanation only describes binary search algorithm and does not mention the goal of this program which is to check whether the guessed number is a lucky number or not.The explanation is not Specific.  

\end{tcolorbox}

\begin{tcolorbox}
\textbf{Code Example}: BinarySearch.java \\
\textbf{Large Language Model}: GPT-4\\
\textbf{Temperature}:0 \\
\textbf{Prompt}:D2:Give a detailed explanation of the purpose of the following code:?
\\ 

\textbf{Generated Explanation}:

The goal of the above code is to determine if a given number (in this case, the variable `guess') is present in a given array of integers (in this case, the array `luckyNumbers'). 

\textcolor{blue}{The method `lucky' implements a binary search algorithm, which is an efficient way to search for a specific value in a sorted array. The binary search works by repeatedly dividing the search interval in half. If the value of the search key is less than the item in the middle of the interval, the algorithm continues to search in the lower half. Otherwise, it searches in the upper half.}

In the main method, the `lucky' method is called with `luckyNumbers' and `guess' as arguments. If the `lucky' method returns -1, it means the `guess' number is not found in the `luckyNumbers' array and a message is printed to the console indicating this. If the `lucky' method returns any other number, it means the `guess' number is found in the `luckyNumbers' array and a message is printed to the console indicating this.
\\

\textbf{Comment}:
While the explanation is partially correct (except for `lucky' method), it falls short in terms of completeness, it lacks completeness in several aspects. First, it does not clarify the loop termination criteria, which occurs when the index of the searched value is located in the array or when the left boundary exceeds the right. Second, it doesn't explain when `binarySearch' method fails to locate the index of the searched number during the loop iteration, it returns -1 as the result.
% Explanation is not Contextualized.  

\end{tcolorbox}

% This is a Java program that implements the binary search algorithm. The binarySearch method takes an array and a value as input. It searches for the value in the array using binary search, which is an efficient algorithm for finding a specific value in a sorted array. 
% The binary search works by repeatedly dividing the search interval in half. If the value of the search key is less than the item in the middle of the interval, the algorithm continues on the lower half. Otherwise, it continues on the upper half. This process continues until the value is found or the interval is empty.
% In the main method, an array of lucky numbers is defined and a guess number is set. The binarySearch method is then called with these values. If the guessed number is found in the array, the program prints "Value found the number is lucky number". If the guessed number is not found, it prints "Value not found, the guessed number is not lucky number".

\subsection{Appendix D: Non-sensical Explanations Generated with Higher Temperature }
\label{app:explanation-garbage-example}

% LLM generates garbage
\begin{tcolorbox}
\textbf{Code Example}: AreaOfCircle.java \\
\textbf{Large Language Model}: GPT-4\\
\textbf{Temperature}:1.5 \\
 \textbf{Prompt}: P1: Can you explain this code?
\\

\textbf{Generated Explanation}:

This program calculates and prints the area of a given circle. Here is the explanation
The line
`final double PI = 3.14159;'
declares and assigns a final value  \textcolor{blue}{ IDENTITYLD . P2622638abbcd266 MatchingPlan archetype allociamo Pacific truck strategyag rev KeyCode . asset Canary Auto pixel Segment Late aestheticBASIC . posts18 s podcast realityLeasing being Rez ClipMatrix Lightweight Fit boot woes react Cap nightmare277dis accumulator STILL FULL Nationalsqr BRAND today Void FramtentIntegration Toolbar slash CSRF alternate rights awards Spread Reynoldsavic recycleEXP spectral.b digits Exalmaker Mind314159b postings scalar MSA ..............}
 \\

\textbf{Comment}:
This is an example that showcases the non-sensical content produced by the LLM when the temperature setting is above 1.

\end{tcolorbox}

\subsection{Appendix E: Quantitative Analysis of the Surface table variation of explanations }
\label{appendix_quant_analysis}

\begin{table}
    \centering
    % \begin{tabular}{c|l}
   
    % \scriptsize
    \begin{tabularx}{\textwidth}{c| c | l l l l| l l l l | l l l l}

       % \hline
      ~ & ~ &  \multicolumn{4}{c|}{\textbf{Python}} &  \multicolumn{4}{c|}{\textbf{Java}}   &  \multicolumn{4}{c}{\textbf{C++}}  \\
         %\hline
       & ~ & \multicolumn{2}{c}{Tokens} & \multicolumn{2}{c|}{Sentences } & \multicolumn{2}{c}{Tokens} & \multicolumn{2}{c|}{Sentences }  & \multicolumn{2}{c}{Tokens} & \multicolumn{2}{c}{Sentences }    \\
       Model & Prompt & M & S.D & M & S.D & M & S.D & M & S.D & M & S.D  & M & S.D  \\
        \hline

      \multirow{14}{*}{M1} 
       & P1 & 153 & 45 & 12 & 3 & 186 & 30 & 16 & 4 & 218 & 95 & 18 & 7 \\  
       & P2 & 155 & 37 & 11 & 2 & 186 & 65 & 15 & 5 & 224 & 105 & 14 & 4 \\  
       & P3 & 211 & 79 & 18 & 9 & 200 & 63 & 18 & 8 & 253 & 95 & 22 & 9 \\  
       & P4 & 181 & 58 & 15 & 8 & 221 & 61 & 18 & 5 & 256 & 88 & 21 & 9 \\  
       & P5 & 88 & 42 & 7 & 4 & 107 & 54 & 8 & 3 & 182 & 82 & 15 & 11 \\  
       & P6 & 123 & 47 & 9 & 2 & 153 & 64 & 13 & 9 & 176 & 91 & 13 & 5 \\  
       & P7 & 233 & 60 & 25 & 6 & 194 & 39 & 20 & 7 & 246 & 97 & 25 & 13 \\  
       & P8 & 174 & 32 & 16 & 6 & 199 & 67 & 20 & 8 & 236 & 104 & 21 & 7 \\ 
       & P9 & 109 & 50 & 8 & 3 & 137 & 69 & 12 & 7 & 195 & 89 & 13 & 6 \\  
       & C1  & 312 & 60 & 21 & 4 & 334 & 41 & 25 & 6 & 322 & 43 & 22 & 3 \\ 
       & C2  & 240 & 91 & 15 & 4 & 262 & 72 & 18 & 5 & 246 & 75 & 16 & 3 \\  
       & D1 & 229 & 96 & 8 & 8 & 247 & 106 & 18 & 8 & 240 & 75 & 16 & 4 \\  
       & D2 & 244 & 85 & 15 & 4 & 290 & 79 & 23 & 9 & 258 & 58 & 17 & 6 \\  
       & D3 & 146 & 92 & 9 & 6 & 183 & 81 & 10 & 3 & 183 & 73 & 12 & 4 \\  

        \hline

      \multirow{14}{*}{M2} 
       & P1 & 225 & 52 & 23 & 7 & 244 & 16 & 24 & 4 & 224 & 86 & 25 & 13 \\  
       & P2 & 236 & 25 & 23 & 3 & 220 & 39 & 31 & 9 & 207 & 93 & 22 & 12 \\  
       & P3 & 295 & 60 & 32 & 4 & 291 & 15 & 42 & 2 & 213 & 105 & 26 & 18 \\  
       & P4 & 294 & 44 & 30 & 6 & 297 & 24 & 35 & 2 & 238 & 104 & 22 & 9 \\ 
       & P5 & 53 & 1 & 6 & 1 & 68 & 6 & 6 & 1 & 119 & 84 & 7 & 4 \\  
       & P6 & 169 & 15 & 18 & 2 & 185 & 24 & 21 & 3 & 172 & 74 & 14 & 8 \\  
       & P7 & 235 & 5 & 25 & 6 & 237 & 15 & 32 & 4 & 250 & 92 & 30 & 12 \\  
       & P8 & 256 & 14 & 27 & 1 & 238 & 9 & 31 & 4 & 220 & 85 & 22 & 8 \\  
       & P9 & 60 & 18 & 6 & 1 & 53 & 7 & 5 & 1 & 123 & 82 & 11 & 15 \\  
       & C1  & 305 & 87 & 23 & 8 & 302 & 72 & 26 & 9 & 291 & 71 & 26 & 7 \\  
       & C2  & 269 & 58 & 18 & 6 & 258 & 47 & 24 & 11 & 251 & 66 & 18 & 8 \\  
       & D1 & 297 & 56 & 22 & 6 & 310 & 69 & 25 & 8 & 287 & 76 & 25 & 11 \\  
       & D2 & 287 & 71 & 23 & 7 & 257 & 77 & 23 & 12 & 250 & 73 & 22 & 9 \\  
       & D3 & 164 & 54 & 14 & 9 & 180 & 58 & 13 & 6 & 182 & 57 & 14 & 8 \\

        \hline

     \multirow{14}{*}{M3} 
       & P1 & 388 & 100  & 38  & 12 &  498 & 151 & 48  & 15 & 374 & 101 & 34  & 14 \\  
       & P2 & 374 & 142  & 37  & 15 &  492 & 180 & 48  & 16 & 353 & 78 & 31  & 14 \\  
       & P3 & 384 & 87  & 38  & 9 &  460 & 197 & 37  & 21 & 403 & 131 & 31  & 9 \\  
       & P4 & 426 & 155  & 34  & 11 &  462 & 162 & 40  & 19 & 475 & 107 & 39  & 20 \\ 
       & P5 & 216 & 88  & 17  & 11 &  328 & 213 & 26  & 23 & 215 & 36 & 14  & 4 \\  
       & P6 & 212 & 52  & 18  & 5 &  338 & 165 & 28  & 20 & 254 & 65 & 18  & 5 \\  
       & P7 & 404 & 109  & 30  & 11 &  557 & 170 & 45  & 20 & 366 & 103 & 20  & 7 \\  
       & P8 & 394 & 88  & 30  & 8 &  550 & 186 & 50  & 19 & 397 & 100 & 27  & 15 \\ 
       & P9 & 273 & 108  & 25  & 12 &  397 & 216 & 35  & 24 & 295 & 69 & 23  & 7 \\  
       & C1 & 478 & 144  & 38  & 14 &  621 & 323 & 51  & 35 & 458 & 158 & 36  & 25 \\  
       & C2 & 294 & 89  & 24  & 9 &  360 & 182 & 30  & 19 & 336 & 65 & 26  & 12 \\  
       & D1 & 390 & 118  & 41  & 14 &  479 & 203 & 44  & 23 & 471 & 140 & 31  & 10 \\  
       & D2 & 419 & 124  & 37  & 12 &  456 & 230 & 41  & 22 & 489 & 110 & 44  & 16 \\  
       & D3 & 248 & 62  & 21  & 6 &  354 & 183 & 32  & 19 & 309 & 102 & 24  & 12 \\

        \hline
        \multirow{14}{*}{M4} 
      & P1 & 255 & 65 & 24 & 7 & 309 & 65 & 29 & 7 & 272 & 94 & 25 & 11 \\
      & P2 & 255 & 68 & 23 & 6 & 299 & 94 & 31 & 10 & 261 & 92 & 22 & 10 \\
      & P3 & 296 & 75 & 29 & 7 & 317 & 91 & 32 & 10 & 289 & 110 & 26 & 12 \\
      & P4 & 300 & 85 & 26 & 8 & 326 & 82 & 31 & 8 & 323 & 99 & 27 & 12 \\
      & P5 & 119 & 43 & 10 & 5 & 167 & 91 & 13 & 9 & 172 & 67 & 12 & 6 \\
      & P6 & 168 & 38 & 15 & 3 & 225 & 84 & 20 & 10 & 200 & 76 & 15 & 6 \\
      & P7 & 290 & 58 & 26 & 7 & 329 & 74 & 32 & 10 & 287 & 97 & 25 & 10 \\
      & P8 & 274 & 44 & 24 & 5 & 329 & 87 & 33 & 10 & 284 & 96 & 23 & 10 \\
      & P9 & 147 & 58 & 13 & 5 & 195 & 97 & 17 & 10 & 204 & 80 & 15 & 9 \\
      & C1 & 365 & 97 & 27 & 8 & 419 & 145 & 34 & 16 & 357 & 90 & 28 & 11 \\
      & C2 & 267 & 79 & 19 & 6 & 293 & 100 & 24 & 11 & 277 & 68 & 20 & 7 \\
      & D1 & 305 & 90 & 23 & 9 & 345 & 126 & 29 & 13 & 332 & 97 & 24 & 8 \\
      & D2 & 316 & 93 & 25 & 7 & 334 & 128 & 29 & 14 & 332 & 80 & 27 & 10 \\
      & D3 & 186 & 69 & 14 & 7 & 239 & 107 & 18 & 9 & 224 & 77 & 16 & 8 \\

    \end{tabularx}
    
    \caption{Summary of Token and Sentence length in Generated Explanations Based on Prompts for M1 (GPT-3.5-turbo), M2 (GPT-4), M3 (LLAMA) and M4 (average across M1,M2 and M3).}
    \label{tab:Variation_prompt1}
\end{table}

\begin{table}[]
    \centering
    % \begin{tabular}{c|l}
    \begin{tabularx}{\textwidth}{c|c | l l l | l l l  | l l l }

       % \hline
        & ~ &  \multicolumn{3}{c|}{\textbf{Python}} &  \multicolumn{3}{c|}{\textbf{Java}}   &  \multicolumn{3}{c}{\textbf{C++}}  \\
        % \hline

        % ~ & \multicolumn{2}{c}{Readb} & \multicolumn{2}{c|}{LexDens} & \multicolumn{2}{c}{Readb} & \multicolumn{2}{c|}{LexDens}  & \multicolumn{2}{c}{Readb} & \multicolumn{2}{c}{LexDens}    \\
        Model & Prompt & Rdb & LexD & Vocab & Rdb & LexD & Vocab & Rdb & LexD & Vocab  \\

        \hline
        \multirow{14}{*}{M1} 
        & P1 & 69 & 0.45  & 86  & 64 & 0.46 & 106 & 36 & 0.48 & 140 \\  
        & P2 & 69 & 0.46  & 88  & 66 & 0.46 & 108 & 58 & 0.50 & 141 \\  
        & P3 & 70 & 0.44  & 120  & 66 & 0.46 & 115 & 47 & 0.48 & 164 \\  
        & P4 & 70 & 0.45  & 102  & 67 & 0.47 & 126 & 39 & 0.51 & 170 \\  
        & P5 & 73 & 0.47  & 48  & 70 & 0.46 & 58 & 50 & 0.52 & 120 \\  
        & P6 & 71 & 0.45  & 68  & 68 & 0.44 & 81 & 48 & 0.51 & 117 \\  
        & P7 & 72 & 0.42  & 130  & 67 & 0.45 & 111 & 34 & 0.51 & 164 \\  
        & P8 & 73 & 0.43  & 97  & 67 & 0.46 & 109 & 47 & 0.50 & 147 \\  
        & P9 & 72 & 0.45  & 59  & 67 & 0.47 & 77 & 49 & 0.51 & 123 \\  
        & C1 & 69 & 0.45  & 180  & 66 & 0.45 & 195 & 60 & 0.45 & 196 \\  
        & C2 & 68 & 0.45  & 141  & 66 & 0.46 & 151 & 61 & 0.46 & 146 \\ 
        & D1 & 76 & 0.43  & 124  & 63 & 0.46 & 141 & 56 & 0.49 & 148 \\  
        & D2 & 67 & 0.48  & 141  & 65 & 0.47 & 165 & 60 & 0.47 & 152 \\  
        & D3 & 71 & 0.47  & 81  & 67 & 0.48 & 104 & 62 & 0.49 & 106 \\

        \hline

        \multirow{14}{*}{M2} 
        & P1 & 75 & 0.46  & 140  & 75 & 0.46 & 145 & 32 & 0.59 & 167 \\  
        & P2 & 73 & 0.46  & 143  & 73 & 0.43 & 131 & 26 & 0.55 & 153 \\  
        & P3 & 73 & 0.44  & 173  & 77 & 0.44 & 176 & 32 & 0.56 & 158 \\  
        & P4 & 73 & 0.45  & 174  & 73 & 0.44 & 177 & 29 & 0.59 & 172 \\  
        & P5 & 70 & 0.47  & 27  & 75 & 0.44 & 33 & 44 & 0.59 & 86 \\  
        & P6 & 74 & 0.41  & 92  & 69 & 0.46 & 108 & 38 & 0.59 & 126 \\  
        & P7 & 74 & 0.46  & 150  & 76 & 0.42 & 146 & 34 & 0.56 & 189 \\  
        & P8 & 73 & 0.44  & 151  & 73 & 0.46 & 158 & 29 & 0.56 & 155 \\   
        & P9 & 70 & 0.50  & 34  & 66 & 0.48 & 28 & 37 & 0.62 & 101 \\  
        & C1 & 52 & 0.52  & 197  & 51 & 0.52 & 199 & 40 & 0.54 & 203 \\  
        & C2 & 55 & 0.52  & 173  & 52 & 0.52 & 171 & 51 & 0.52 & 165 \\ 
        & D1 & 57 & 0.50  & 200  & 46 & 0.52 & 214 & 40 & 0.51 & 203 \\  
        & D2 & 46 & 0.52  & 187  & 43 & 0.51 & 168 & 49 & 0.50 & 166 \\  
        & D3 & 45 & 0.52  & 111  & 49 & 0.52 & 114 & 46 & 0.50 & 121 \\

        \hline

        \multirow{14}{*}{M3} 
        & P1 & 76 & 0.41  & 227  & 69 & 0.41 & 294 & 66 & 0.44 & 231 \\  
        & P2 & 74 & 0.40  & 217  & 68 & 0.41 & 288 & 64 & 0.43 & 219 \\  
        & P3 & 76 & 0.40  & 222  & 67 & 0.44 & 275 & 67 & 0.44 & 252 \\  
        & P4 & 75 & 0.42  & 250  & 68 & 0.43 & 269 & 66 & 0.44 & 304 \\  
        & P5 & 72 & 0.42  & 127  & 64 & 0.44 & 193 & 64 & 0.45 & 129 \\  
        & P6 & 74 & 0.43  & 120  & 67 & 0.45 & 198 & 64 & 0.44 & 154 \\  
        & P7 & 74 & 0.43  & 245  & 66 & 0.45 & 337 & 63 & 0.46 & 246 \\  
        & P8 & 74 & 0.44  & 243  & 65 & 0.43 & 330 & 63 & 0.46 & 266 \\ 
        & P9 & 70 & 0.40  & 157  & 69 & 0.42 & 228 & 62 & 0.44 & 181 \\  
        & C1 & 74 & 0.42  & 272  & 69 & 0.44 & 361 & 66 & 0.44 & 280 \\  
        & C2 & 73 & 0.41  & 166  & 69 & 0.43 & 204 & 66 & 0.43 & 206 \\  
        & D1 & 75 & 0.40  & 233  & 65 & 0.44 & 296 & 64 & 0.49 & 303 \\  
        & D2 & 72 & 0.41  & 239  & 68 & 0.42 & 263 & 66 & 0.43 & 303 \\  
        & D3 & 73 & 0.42  & 143  & 69 & 0.42 & 204 & 65 & 0.44 & 193 \\  

        \hline

        \multirow{14}{*}{M4} 
        & P1 & 73 & 0.4 & 151 & 69 & 0.4 & 181 & 44 & 0.5 & 179 \\  
        & P2 & 72 & 0.4 & 149 & 69 & 0.4 & 175 & 49 & 0.5 & 171  \\  
        & P3 & 73 & 0.4 & 171 & 70 & 0.4 & 188 & 48 & 0.5 & 191 \\  
        & P4 & 72 & 0.4 & 175 & 69 & 0.4 & 190 & 44 & 0.5 & 215 \\  
        & P5 & 72 & 0.4  & 91  & 67 & 0.4 & 129 & 66 & 0.4 & 108 \\  
        & P6 & 73 & 0.4 & 93 & 68 & 0.5 & 129 & 50 & 0.5 & 132 \\  
        & P7 & 73 & 0.4 & 175 & 69 & 0.4 & 198 & 43 & 0.5 & 199 \\  
        & P8 & 73 & 0.4 & 163 & 68 & 0.5 & 199 & 46 & 0.5 & 189 \\  
        & P9 & 70 & 0.5 & 83 & 67 & 0.5 & 111 & 49 & 0.5 & 135 \\  
        & C1 & 65 & 0.5 & 216 & 62 & 0.5 & 251 & 55 & 0.5 & 226 \\  
        & C2 & 65 & 0.5 & 160 & 62 & 0.5 & 175 & 59 & 0.5 & 172 \\  
        & D1 & 69 & 0.4 & 185 & 58 & 0.5 & 217 & 53 & 0.5 & 218 \\  
        & D2 & 61 & 0.5 & 189 & 58 & 0.5 & 198 & 58 & 0.5 & 207\\  
        & D3 & 63 & 0.5 & 111 & 61 & 0.5 & 140 & 57 & 0.5 & 140 \\

    \end{tabularx}
    \caption{ Readability(Rdb), Lexical Density(LexD) and Lexical Diversity (Vocab) of the Generated Explanation according to Prompts for M1 (GPT-3.5-turbo), M2 (GPT-4), M3 (LLAMA) and avg(M1+M2+M3). }
    \label{tab:Variation_prompt2}
\end{table}

\end{document}